# Degenerate and non-degenerate parametric excitation in YIG nanostructures

H. Merbouche[1*], P. Che[2], T. Srivastava[3], N. Beaulieu[4], J. Ben Youssef[4], M. Muñoz[5], M. d'Aquino[6], C. Serpico[6], G. de Loubens[3], P. Bortolotti[2], A. Anane[2], S. O. Demokritov[1], and V. E. Demidov[1]

[1]*Institute for Applied Physics, University of Muenster, Muenster, Germany*

[2]*Unité Mixte de Physique, CNRS, Thales, Université Paris-Saclay, Palaiseau, France*

[3]*SPEC, CEA, CNRS, Université Paris-Saclay, Gif-sur-Yvette, France*

[4]*LabSTICC, UMR 6285 CNRS, Université de Bretagne Occidentale, Brest, France*

[5]*Instituto de Tecnologías Físicas y de la Información (CSIC), Madrid, Spain*

[6]*Department of Electrical Engineering and ICT, University of Naples Federico II, Italy*

We study experimentally the processes of parametric excitation in microscopic magnetically saturated disks of nanometer-thick Yttrium Iron Garnet. We show that, depending on the relative orientation between the parametric pumping field and the static magnetization, excitation of either degenerate or non-degenerate magnon pairs is possible. In the latter case, which is particularly important for applications associated with the realization of computation in the reciprocal space, a single-frequency pumping can generate pairs of magnons whose frequencies correspond to different eigenmodes of the disk. We show that, depending on the size of the disk and the modes involved, the frequency difference in a pair can vary in the range 0.1-0.8 GHz. We demonstrate that in this system, one can easily realize a practically important situation where several magnon pairs share the same mode. We also observe the simultaneous generation of up to six different modes using a fixed-frequency monochromatic pumping. Our experimental findings are supported by numerical calculations that allow us to unambiguously identify the excited modes. Our results open new possibilities for the implementation of reciprocal-space computing making use of low-damping magnetic insulators.

*Corresponding author, e-mail: Hugo.MERBOUCHE@uni-muenster.de

# I. INTRODUCTION

The field of magnonics aims to use spin waves (SWs) as information carriers to perform radio-frequency (rf) processing as well as classical and neuromorphic computing [1–4]. Indeed, spin-waves (and their quanta – magnons) represent a promising alternative to charge carriers thanks to their low energy, short wavelength, and easily attainable nonlinear dynamics. These characteristics make them particularly suited for the implementation of new computing schemes that take advantage of the massive parallelization of operations in the frequency space and of the intrinsic hyperconnectivity in the reciprocal space (k-space) due to non-linear spin-wave interactions [5–11]. Dynamic magnetic systems can thus be seen as a neural network whose nodes are SW modes that are interconnected by nonlinear magnon-magnon interactions in the k-space. The inputs and outputs of such network can be either accessed and read in the frequency space [10,11], or by employing a real-space separation, through propagation and wave interference [5,6,8]. Implementing such schemes requires the excitation and manipulation of many modes in small magnetic structures. This can be achieved by using parametric processes [12], where a photon or a magnon at frequency $2f$ splits in two magnons at frequencies $f - \delta f$ and $f + \delta f$. The splitting can be degenerate ($\delta f = 0$) or non-degenerate ($\delta f \neq 0$). The degenerate process has been the most studied to excite SW modes in a broad range of materials and systems including extended films [12–18] and micro- and nano-waveguides [19–22] as well as magnetic nanocontacts [23], magnetic tunnel junctions [24], micro- and nano-dots of Permalloy [25–27] and Yttrium Iron Garnet [28] (YIG). Contrary to direct linear excitation [29], it allows the efficient excitation of modes using a uniform dynamic magnetic field irrespectively of the mode spatial profile [25,28]. On the other hand, excitation of non-degenerate pairs have been observed only in extended micrometer-thick YIG films [30] and in confined metallic structure in a vortex state [31,32]. In the latter case, the non-degeneracy was shown to open the possibility to cross-stimulate mode excitation using multiple parametric pumps, thus effectively implementing an interconnected recurrent neural network capable of classifying microwave signals [11].

So far, the non-degenerate parametric excitation relied on the specific configuration with a strong dipolar dip in the dispersion relation, which was achieved in either micrometer-thick YIG [30] films or in ferromagnetic structures in the vortex ground state at zero [31] or low static magnetic fields [32]. In these configurations, the dispersion relation is such that the frequencies of the low-order magnon modes are about twice the frequencies of higher-order modes. The low-order modes can be efficiently excited using direct inductive mechanism and then generate magnon pairs through the parametric process [30,33]. Such favorable configurations are not possible in nanometer-thick films of magnetic insulators with low magnetization, such as YIG. Yet, YIG is a key material in magnonics due to its extremely low magnetic damping $\alpha$ [34–36], nearly two orders of magnitude smaller than that in metallic ferromagnets. Since low damping strongly lowers the threshold of nonlinear processes [12,33], finding a method to excite non-degenerate magnon pairs in ultrathin-film YIG structures is an important step in the development of magnonic computing schemes based on nonlinear interactions.

In this study, we demonstrate experimentally that both degenerate or non-degenerate parametric processes can be realized in the same magnetically saturated microscopic ultrathin YIG disk by simply varying the direction of the parametric pumping field with respect to the static magnetization. When the field is applied parallel to the static magnetization (parallel pumping), a photon of the pumping field splits into a degenerate magnon pair at half frequency, as expected. However, when the field is applied transverse to the static magnetization (transverse pumping), it non-resonantly excites a magnon which then splits into a magnon pair that is typically non-degenerate. We show that the efficiency of this process is comparable to the parallel-pumping process, which makes it attractive for technical application. Additionally, the non-degenerate parametric excitation via the generation of non-resonant magnons is not limited by the frequencies of the resonant modes of the system. As a result this technique is much more flexible compared to resonant 3-magnon scattering schemes. These findings greatly facilitate the implementation of promising k-space computing schemes using attractive magnonic materials such as YIG.

## II. EXPERIMENT

Figure 1(a) shows the schematics of the experiment. A 52-nm thick YIG film grown by liquid-phase epitaxy is patterned using e-beam lithography to define disks with a diameter $D$ ranging from 500 nm to 2 µm. A 150-nm-thick and 4-µm wide Au antenna is fabricated on top of the YIG structures to apply a uniform dynamic magnetic field $h_{rf}$ induced by the rf current $i_{rf}$ in the antenna. The static magnetic field $H_0$ is applied in-plane such that it is either transverse ($\varphi = 90°$) or parallel ($\varphi = 0$) to the dynamic field $h_{rf}$. The excitation is applied in the form of 500-ns-long pulses with the repetition period of 1 µs to monitor the transient dynamics of the different parametric processes. The magnetization dynamics is detected using micro-focus Brillouin Light Scattering (BLS) spectroscopy [37], which yields a signal (BLS intensity) proportional to the intensity of magnetization dynamics at the position of the probing laser spot (see Fig. 1(a)). The BLS measurements are performed with simultaneous spectral and temporal resolution. The latter is realized by synchronizing the detection of the scattered light with the excitation pulses.

We first characterize the spectrum of resonant SW modes by using a direct low-amplitude linear excitation. We apply the static field transverse to the dynamic field ($\varphi = 90°$), vary the frequency $f$ of the dynamic field in a broad range and record the rf-induced BLS intensity for each excitation frequency. The obtained resonant spectrum for the 500-nm disk at $H_0 = 20$ mT and $P = 0.2$ mW is plotted in the logarithmic scale in Fig. 1(b). Despite the poor coupling of the uniform excitation field with non-uniform spin-wave modes, up to 8 peaks can be detected in the spectrum. To identify the observed modes, we calculate the frequencies and spatial profiles of the eigenmodes of the disk using an eigenmode solver [38]. In these calculations we use the independently determined saturation magnetization 176 mT and standard for YIG exchange constant 3.6 pJ/m. Due to the small diameter of the disk, the modes are well separated in the frequency space. Therefore, it is easy to associate the frequencies obtained from calculations (vertical lines in Fig. 1(b)) with the frequencies of the peaks observed in the experiment. All the computed mode frequencies match well with the measured frequencies, except for the first two lowest-frequency modes. As seen from the computed mode profiles in Fig. 1(b), these are the modes strongly localized at the edges of the disk – edge modes. We note that this is a generic feature of edge modes to be very sensitive to the conditions at the edges [39]. As a result, their quantitative characteristics typically cannot be reproduced in calculations implying ideal edges [40–42]. Nevertheless, from the comparison of the experimental spectrum with the calculated one, it is natural to associate the two peaks at frequencies 1.30 and 1.36 GHz to the edge modes of the disk. In contrast, the identification of the bulk modes is straightforward. These modes can be labelled using the number of lobes $(n_\parallel, n_\perp)$ of the standing waves in the directions parallel and transverse to the in-plane static magnetic field $H_0$. The first three calculated modes of the $n_\perp = 1$ branch have frequencies $f_{1,1}=1.63$ GHz, $f_{2,1}=1.67$ GHz, and $f_{3,1}=1.91$ GHz, which agree well with the frequencies of the observed peaks 1.64, 1.70, and 1.94 GHz. The calculated frequency of the mode (1,2) $f_{1,2}=2.19$ GHz belonging to the branch $n_\perp = 2$ coincides with the detected frequency of the peak at 2.19 GHz, and the frequencies of the modes (1,3) $f_{1,3}$=2.59 GHz and (2,3) $f_{2,3}$=2.55 GHz belonging to the branch $n_\perp = 3$ fit the frequencies of the BLS peaks at 2.61 GHz and 2.55 GHz.

## III. RESULTS AND DISCUSSION

We now investigate the parametric excitation of the modes using high-amplitude dynamic fields. First, we address the well-studied case of the parallel pumping corresponding to the geometry where the static magnetic field is oriented parallel to the dynamic excitation field $h_{rf}$ ($\varphi = 0$). Under these conditions, no direct linear coupling of the excitation field with the dynamic magnetization is possible. Instead, the magnetization dynamics can be driven by the parametric mechanism requiring the frequency of the excitation field to be twice the frequency of the modes to be excited. Figure 2(a) shows a representative BLS spectrum of magnetic oscillations recorded by applying the rf current with a power $P = 2$ mW at the pumping frequency $f_p$=3.28 GHz, which is twice the frequency of the fundamental (1,1) mode of the disk 1.64 GHz (see Fig. 1(b)). As seen from Fig. 2(a), in agreement with the expectations, pumping efficiently excites the mode at $f_p/2$. Figure 2(b) combines in a color map the BLS spectra measured for pumping frequencies varying from 2.4 to 4.8 GHz. These data clearly show that parallel pumping results in the excitation of degenerate magnon pairs at exactly $f_p/2$ (dashed white line), when this frequency coincides with the frequency of one of the resonant modes of the disk. Since the parametric process is not limited by the spatial profile of the excitation field, many more modes can be excited in comparison with the case of linear excitation (Fig. 1(b)). Additionally, since the parametric process is a threshold phenomenon, the number of excited modes depends on the pumping power. This is demonstrated in Fig. 2(c), which shows the power dependence of the intensity of the modes labelled as A, B, and C in Fig. 2(b). In particular, the mode C can only be excited at $P > 1$ mW, while the mode A exhibits non-zero intensity already at $P = 0.05$ mW. The threshold power is known to be proportional to the ratio of the relaxation frequency to the ellipticity [12,27]. As the relaxation frequency increases and the ellipticity of the modes tends to decrease with increased $n_{\parallel}$ and $n_{\perp}$, the threshold tends to increase with frequency.

We would like to emphasize that although the number of excited modes increases with the pumping power, the process remains degenerate at all powers. In other words, by using pumping with a fixed frequency, only one mode can be excited at a time. A careful examination of the data in Fig. 2(b) shows that, in addition to the signal at $f_p/2$, one also observes a weak signal at $f_p$ (e.g., mode labelled as D). This weak additional excitation appears only if the amplitude of the mode at $f_p/2$ is large and, therefore, is not associated with the direct excitation of magnetization dynamics by the pumping. Moreover, time-resolved BLS measurements show that the signal at $f_p/2$ precedes the signal at $f_p$ in time (Fig. 2(d)), which is a clear indication that the signal at $f_p$ is caused by the parametrically excited dynamics at $f_p/2$, and not vice versa. We associate this signal with a non-resonant second harmonic generation [43].

Let us now return to the configuration of transverse pumping ($\varphi = 90°$), which we used earlier for linear excitation (Fig. 1(b)) and apply an excitation field at frequencies twice the frequencies of the detected modes. Under these conditions, the excitation field can linearly excite magnetization dynamics. However, the resonant modes available at these frequencies possess very large effective wavevectors and do not couple efficiently to the uniform dynamic field. Therefore, the non-resonant excitation dominates in this regime. At sufficiently large powers, the non-resonantly excited magnon at $f_p$ can split into two magnons in the same way as it happens in the case of the interaction of resonant magnons [30–32]. Moreover, as in the case of resonant magnons, this parametric process is expected to be non-degenerate.

Figure 3(a) shows a representative BLS spectrum recorded at $f_P = 3.48$ GHz and $P = 5$ mW. In contrast to the spectrum for the case of parallel pumping (Fig. 2(a)), the spectrum in Fig. 3(a) exhibits a well-pronounced peak at $f_P$ (labelled B) reflecting the non-resonant excitation of magnetization dynamics at this frequency. As clearly seen from Fig. 3(a), non-resonantly excited magnons B do not split into magnons at $f_P/2$. Instead, a non-degenerate magnon-pair (A1, A2) at $f_P/2 \pm \delta f$ is excited, where $\delta f = 0.44$ GHz. The frequencies of the excited modes A1 (1.3 GHz) and A2 (2.18 GHz) correspond to previously identified first edge mode and the (1,2) mode. Figure 3(b) combines in a color map the BLS spectra measured for pumping frequency varying from 2.8 to 4.8 GHz. As can be seen from these data, in the geometry of transverse pumping, parametric excitation has a non-degenerate character for all the observed splitting processes: there is no signature of dynamics excited at frequency $f = f_P/2$. Additionally, in accordance with expectations, we observe excitation at $f = f_P$ over the entire studied range, which is in strong contrast to the case of parallel pumping (Fig. 2(b)).

To get better insight into the non-degenerate splitting possesses, we mark in Fig. 3(b) the experimentally resolved frequencies of the modes (Fig. 1(b)). From this analysis one can observe that many observed mode pairs include at least one edge mode. This can be associated with their strong spatial localization resulting in a wide distribution in the reciprocal space, which helps to fulfill the linear-momentum conservation law in magnon splitting processes [44]. We emphasize, however, that the eigenmodes of an in-plane saturated magnetic disk are far from being a simple combination of standing plane waves (see mode profiles in Fig. 1(b)). Therefore, even bulk modes are characterized by a wide distribution in the momentum space, which facilitates the fulfilment of the momentum conservation [33,44]. As will be shown below, in larger disks, where the mode spectrum is much denser in frequency space, pairs of bulk modes can also be easily excited.

In order to compare the efficiency of the non-degenerate transverse-pumping process with the parallel-pumping case, we find the threshold power necessary to excite the most intense pair, labelled in Fig. 3(b) as A1 and A2. The power dependence of the intensities of these modes is plotted in Fig. 3(c). We emphasize that the apparent difference in the intensities of the modes forming a pair (see Fig. 3(a)) originates from the different sensitivity of the optical setup to different modes. Therefore, we renormalize the dependence for the mode A2 to match the intensity of the mode A1 (we choose a normalization factor that is slightly off to improve readability). The perfect match between the two curves indicates that both modes are excited simultaneously and possess the same excitation threshold, as expected for a non-degenerate three-magnon process. The value of this threshold 0.4 mW is larger than that for the mode A in Fig. 2(c), which corresponds to the mode A1 of the pair. However, it is more than twice smaller than the threshold of the mode C in Fig. 2(c), which corresponds to the mode A2. In other words, despite the need for non-resonant excitation, the efficiency of the transverse-pumping parametric process is comparable to that of the parallel-pumping process. A possible approach to further improve the efficiency can be based on the use of magnetic systems combing two materials with different saturation magnetization, where transverse pumping resonantly excites magnons in one material, which then parametrically excite magnon pairs in the other material [45].

Let us now turn to the temporal characteristics of the non-degenerate process. Figure 3(d) shows the temporal dependence of the intensities of the modes A1 and A2 together with the dependence of the intensity of oscillations at the pumping frequency $f_\text{p}$ (labelled as B in Fig. 3(b)). First, one can see that rise time of the intensities A1 and A2 until the steady state is reached is somewhat longer than in the case of the parallel-pumping excitation (Fig. 2(d)). Second, in strong contrast to the parallel-pumping case, the signal at $f_\text{p}/2 \pm \delta f$ is delayed with respect to the signal at $f_\text{p}$. This reflects the two-stage nature of the transverse-pumping process, which requires the excitation of an intense non-resonant dynamic magnetic state as an intermediate step. We note, however, that the settling times are of the same order of magnitude. Additionally, they can be reduced by further increasing the pumping power.

We now discuss the effect of the disk size on the excitation of non-degenerate pairs. Figures 4(a) and 4 (b) show the excitation maps for the disks with the diameter $D = 1$ μm and 2 μm, respectively. As seen from these data, more pairs can be excited in larger disks, and most observed pairs belong to the frequency range of bulk modes. This is associated with the decrease in the mode separation in the frequency space with increasing diameter $D$, which also results in a decrease of the characteristic frequency separation in individual pairs $2\delta f$. In particular, for $D = 1$ μm $2\delta f \approx 0.2$ GHz, and for $D = 2$ μm $2\delta f \approx 0.1$ GHz, i.e., the frequency separation seems to be inversely proportional do the disk size. Due to the denser mode spectrum, in larger disks, one can easily find pairs that share the same mode, e.g., pairs A+B and B+C in Fig. 4(a). If such pairs are driven simultaneously by a multi-frequency pumping field, one can expect cross-stimulation of splitting processes, as was previously observed for resonant three-magnon splitting in the vortex state [11]. This opens, for example, the possibility to significantly expand the range of materials and geometries for implementation of non-traditional computing scheme proposed in Ref. [11]. Additionally, many of the modes belonging to the observed pairs can be easily excited using a linear excitation mechanism, which provides an additional means to control the system of interacting modes. Moreover, due to the dense spectrum of modes in sufficiently large disks, it becomes possible to simultaneously fulfill the energy conservation conditions for several pairs. As a result, one can excite multiple pairs using a single-frequency transverse pumping. This is demonstrated in Fig. 4(c), which shows the BLS spectrum recorded for a 2-μm large disk driven by the pumping at $f_\mathrm{P}$=3.82 GHz (also marked with an arrow in Fig. 4(b)). As seen from these data, up to 3 mode pairs (6 different modes) can be simultaneously excited in this system.

## IV. CONCLUSIONS

In conclusion, we have shown that a relatively simple microscopic system based on a promising low-damping magnetic insulator can demonstrate a large variety of parametric phenomena that can be controlled by varying the relative orientation between the pumping field and the static magnetization. In particular, the system allows one to simultaneously excite many dynamical modes by applying a single-frequency excitation signal. This possibility is of decisive importance to implement novel hardware platforms for non-traditional computing and data processing relying on modes interacting in the reciprocal space. Our findings show that practical implementation of such systems is not limited to specific static magnetic configurations and materials but can also be based on versatile low-damping thin magnetic insulators, which is expected to enable further development of the magnonic neuromorphic computing field.

## ACKNOWLEDGMENTS

This work was supported by the Horizon2020 Research Framework Program of the European Commission under grant no. 899646 (k-NET).

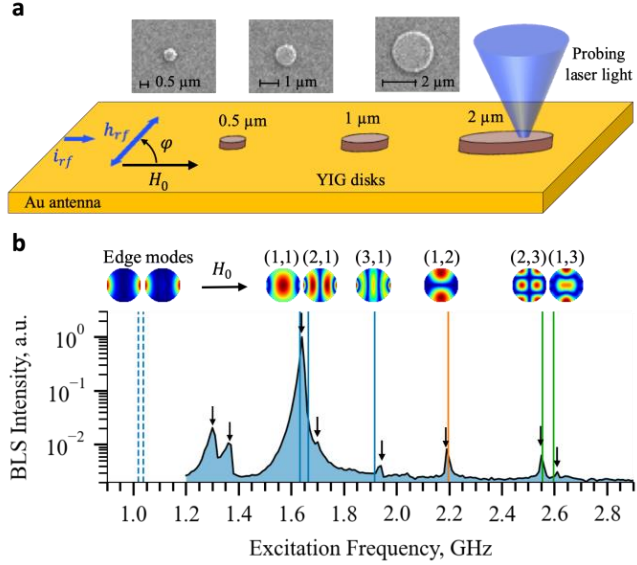

**Figure 1**: (a) Schematics of the experiment. YIG disks with a thickness 52 nm and diameter 0.5, 1 and 2 µm and are excited by a dynamic field $h_{rf}$ created using a 4-µm wide antenna. The static magnetic field $H_0$ is applied either parallel or transverse to $h_{rf}$. Insets show the electron-microscope images of the disks. (b) Characterization of the eigenmode spectrum of 0.5-µm disk. The graph shows the intensity of the dynamic magnetization as a function of the excitation frequency under conditions of direct linear excitation ($\varphi = 90°$) with $P = 0.2$ mW at $H_0 = 20$ mT. Arrows mark the frequencies of the detected resonant peaks. Vertical lines mark the calculated frequencies of the eigenmodes. The spatial profiles of the modes are shown close to the corresponding lines. These profiles present the absolute value of the out-of-plane component of the dynamic magnetization.

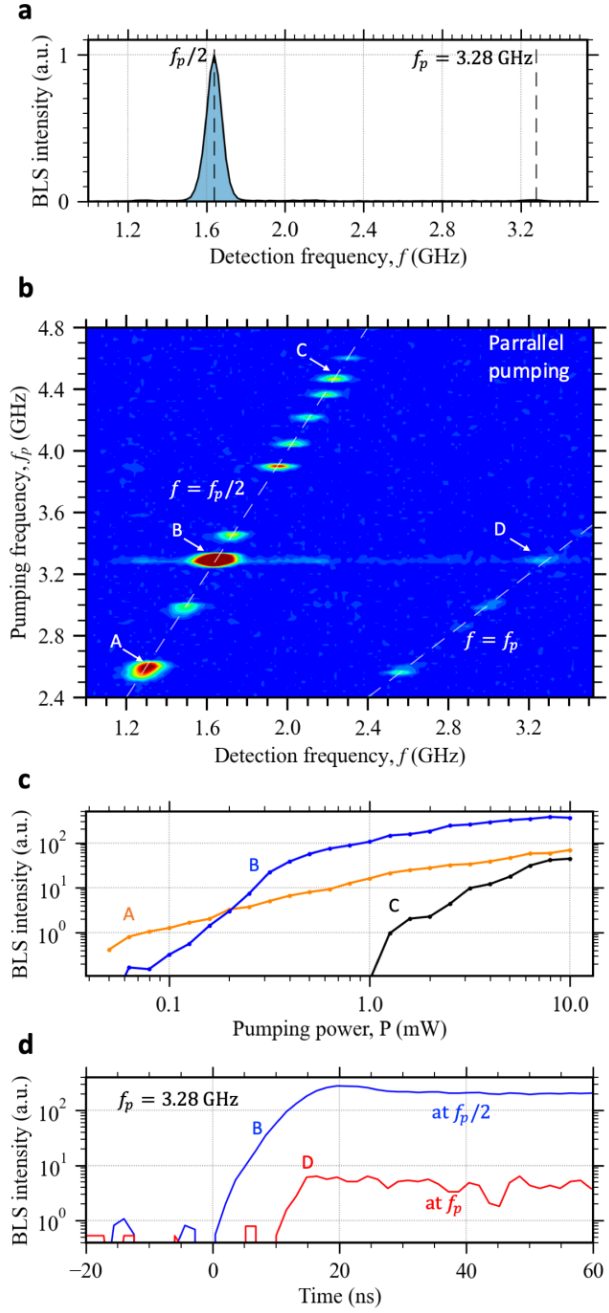

**Figure 2**: (a) Representative BLS spectrum of parametrically-excited magnetic oscillations recorded at the pumping frequency $f_p$=3.28 GHz and pumping power $P = 2$ mW. (b) BLS spectra measured for pumping frequency $f_p$ varying from 2.4 to 4.8 GHz combined in a color map. (c) Power dependences of the intensity of the modes labelled as A, B, and C in (b). (d) Temporal dependences of the modes labelled as B and D in (b) measured at $P = 10$ mW. The data were obtained at $H_0 = 20$ mT and $\varphi = 0°$ (parallel-pumping geometry) on the $D = 0.5$ µm disk.

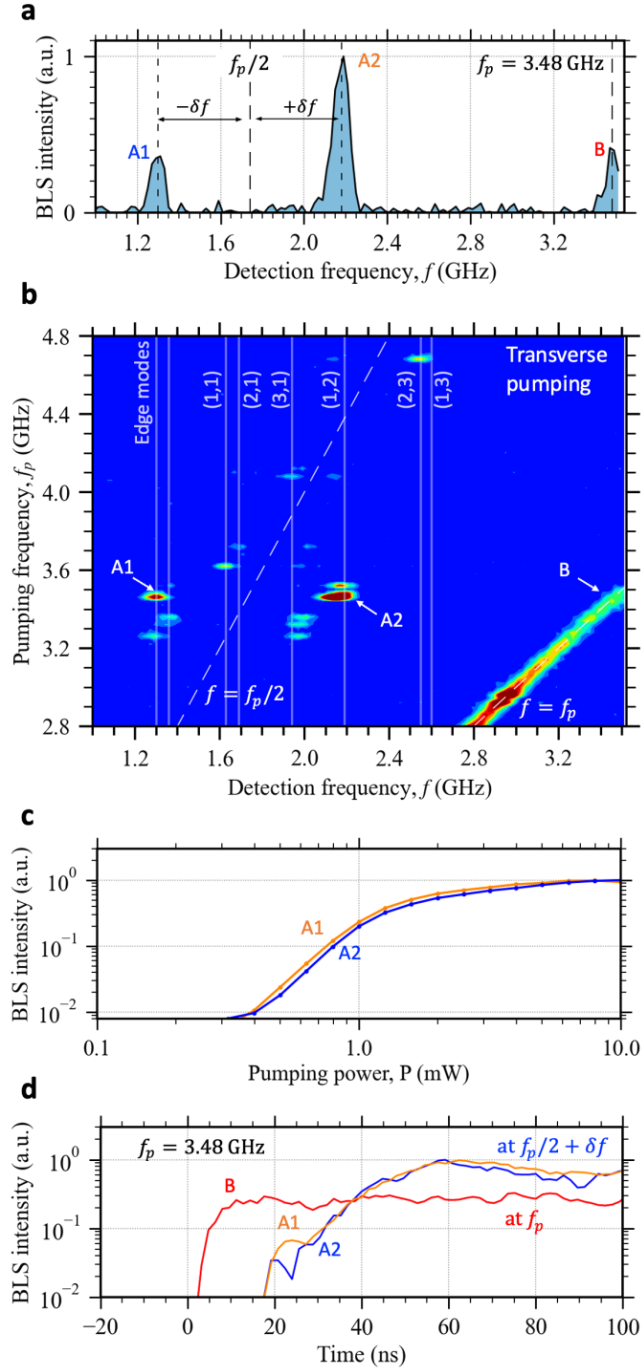

**Figure 3**: (a) Representative BLS spectrum of parametrically-excited magnetic oscillations recorded at the pumping frequency $f_p$=3.48 GHz and pumping power $P = 5$ mW. (b) BLS spectra measured for pumping frequency $f_p$ varying from 2.8 to 4.8 GHz combined in a color map. Vertical lines mark the experimentally-detected frequencies of the eigenmodes. (c) Power dependences of the intensity of the modes labelled as A1 and A2 in (b). (d) Temporal dependences of the modes labelled as A1, A2, and B in (b) measured at $P = 10$ mW. The data were obtained at $H_0 = 20$ mT and $\varphi = 90°$ (transverse-pumping geometry) on the $D = 0.5$ μm disk.

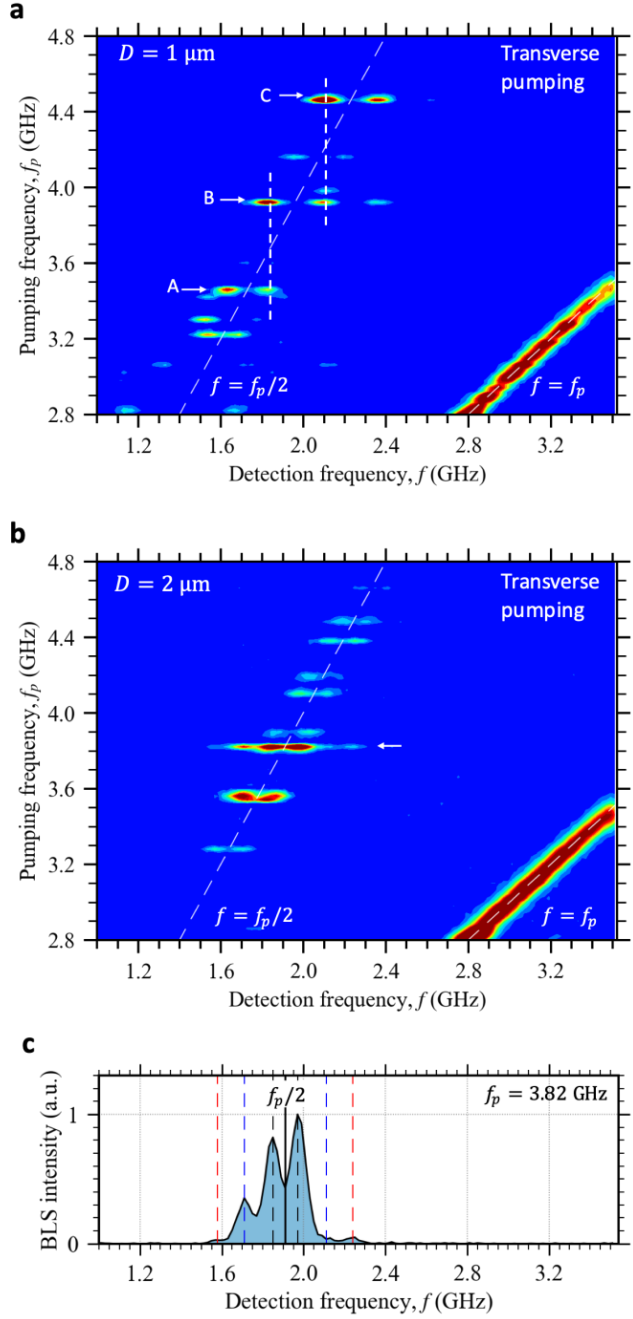

**Figure 4**: (a) and (b) BLS spectra measured for pumping frequency $f_p$ varying from 2.8 to 4.8 GHz combined in a color map. (a) $D = 1$ µm, (b) $D = 2$ µm. Vertical dashed lines in (a) mark the frequencies of the modes common for pairs A, B, and C. (c) BLS spectrum recorded for the 2-µm disk at $f_p$=3.48 GHz, as marked by an arrow in (b). The data were obtained at $P = 5$ mW, $H_0 = 20$ mT and $\varphi = 90°$ (transverse-pumping geometry).